\documentclass[%
 reprint,
 amsmath,amssymb,
 aps, prl
]{revtex4-1}

\usepackage[pdftex]{graphicx}
\usepackage{dcolumn}
\usepackage{bm}
\usepackage[colorlinks=true,linkcolor=blue]{hyperref}
\usepackage[T2A]{fontenc}  
\usepackage[utf8]{inputenc} 
\usepackage[english]{babel} 
\usepackage{xcolor}
\usepackage{enumitem}
\usepackage{subcaption}

\begin{document}

\preprint{PAPER-2024}

\title{Extraordinary manifestation of near electrostatic field caused by macroscopic quantum shell effects in submicron hemispherical clusters}

\author{S.~E.~Kuratov}
\email{ser.evg.kuratov@gmail.com}
\affiliation{Dukhov Research Institute of Automatics, 22 Sushchevskaya st., 127030 Moscow, Russia}

\author{I.~S.~Galtsov}
\affiliation{Dukhov Research Institute of Automatics, 22 Sushchevskaya st., 127030 Moscow, Russia}
\affiliation{Joint Institute for High Temperatures, RAS, 13/2 Izhorskaya st., 125412 Moscow, Russia}

\author{S.~A.~Dyachkov}
\affiliation{Dukhov Research Institute of Automatics, 22 Sushchevskaya st., 127030 Moscow, Russia}
\affiliation{Joint Institute for High Temperatures, RAS, 13/2 Izhorskaya st., 125412 Moscow, Russia}

\author{S.~Yu.~Igashov}
\affiliation{Dukhov Research Institute of Automatics, 22 Sushchevskaya st., 127030 Moscow, Russia}

\date{\today}

\begin{abstract}
  The existence of macroscopic shell structure of submicron metal clusters is known for several decades. Since the most studies provide theoretical analysis for clusters of spherical shape, the electron density inhomogeneities caused by shell effects are spherically symmetric and do not provide long range electrostatic fields. However, similar shell structure should exist in a hemispherical cluster which conserves the closed periodic orbits of electrons, but not the spherical symmetry of electron distribution. As a result, we demonstrate that a strong electrostatic field ($E \sim 10^8$~V/m) exists in the vicinity of the flat surface of an isolated, uncharged metal nanocluster of hemispherical shape using modern approaches for electronic structure evaluation. This physical phenomenon is a consequence of the large-scale spatial inhomogeneity in distribution of electrons related to quantum shell effects in submicron metal clusters, which may find numerous applications in various fields of science and technology. 
\end{abstract}

\maketitle

Metal clusters containing up to thousands atoms are known to manifest macroscopic shell structure of electron subsystem which is observed in various experimental results for ionization potentials, polarizabilities, optical and other properties~\cite{deHeer1993physics}. This is due to the ability of valence electrons to fill the potential well of a whole cluster in the self-consistent manner, which is successfully described using the jellium model for ions with either Hartree-Fock~(HF) or density functional theory~(DFT)~\cite{brack1993physics}. Electrons obeying the Pauli principle fill a finite spatial domain (cluster) with a discrete energy spectrum, which leads to a formation of specific oscillations in radial density distribution of bound electrons. Adding of an atom to such cluster may affect much the density distribution, the self-consistent field, and the binding energy per particle resulting in the sawtooth dependency on the particle number. The particle numbers providing energy minima are also referred to as ``magic numbers''.

Quantum shell effects in a spherical mesoscopic system of degenerate electrons may be described using various theoretical approaches: 
\begin{enumerate}
  \item the jellium model for ions and the quantum mechanical models (HF, DFT, etc.) for electrons;
  \item the direct numerical simulation of the self-consistent field of electrons and ions (as implemented e.g. in VASP~\cite{Kresse:1996} and other well-known DFT packages);
  \item the analytical method of semiclassical Green's functions (trace formula \cite{gutzwiller2013chaos}).
\end{enumerate}
As shown in Ref.~\cite{brack1993physics}, Approach 1 is extensively applied for spherically symmetric clusters, as soon as the problem is reduced to the one-dimensional case. This allows to evaluate some properties of large clusters up to tens of thousands of atoms. Approach 2 in general is related to quantum chemistry, allowing three-dimensional modeling of systems containing up to thousands of particles by means of modern supercomputing. Approach 3 may provide some insight to the properties of the system with rather small numerical cost and even allows to obtain compact analytical expressions by utilizing symmetries of the cluster.

Recently, Approach 3 was applied to the analysis of closed nonperiodic orbits in a spherically symmetric potential for describing oscillating contributions to the spatial distribution of electronic density~\cite{roccia2008closed,roccia2010semiclassical}. It was noted that there are two types of density oscillations: regular ones characterized by a small spatial scale, and irregular ones with a noticeably larger scale, which can arise in a problem with more than one spatial dimension. However, there was quite small attention paid to the analysis of spatial properties these oscillations, while the mathematical aspect of the problem was discussed in details.

\begin{figure*}[t]
\centering
\includegraphics[width=0.99\textwidth]{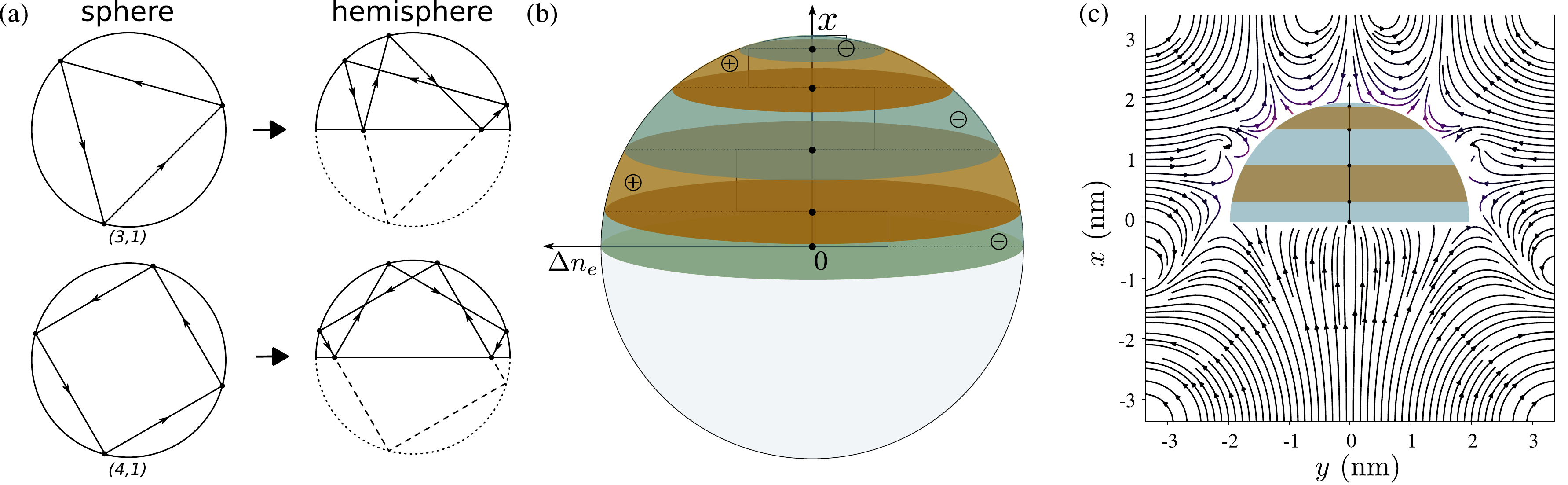}
\caption{(a) Example of closed periodic trajectories transformation from spherical to hemispherical cluster. (b) Scheme of a macroscopic shell structure within a spherical cluster reduced to ``layers model'' in a hemisphere: it contains the radial electron density oscillations, so that the resulting charge of a hemisphere is 0, while the near electrostatic field exists. (c) Electric field evaluated for a hemisphere with ``layers model'' electron density distribution. }
\label{fig:hemisphere-model}
\end{figure*}

In addition to the results discussed above, our previous studies~\cite{Kuratov:2019,Kuratov:2021} reveal a nontrivial manifestation of quantum effects in a spherical mesoscopic system of up to $N_e\sim10^9$ degenerate electrons. An unexpected absence of the simple tendency toward a uniform distribution of the electron density has been found, in contrast to that has been tacitly assuming in the limit of large $N_e$. It has been shown that the electron density has a large spatial scale which is of the order of the system size and is much larger than another spatial scale, namely the Fermi wavelength of the degenerate electron gas. This large scale effect clearly manifests itself in the appearance of an electric field acting on the ionic system. The spatial distribution of the potential has an oscillating behavior with several extrema. This result obtained using the above mentioned approaches are found to be in good qualitative and quantitative agreement.

The phenomenon is related to the manifestation of shell effects and can be intuitively understood using the semiclassical Green's function method, which involves integrating over closed classical trajectories~\cite{gutzwiller2013chaos}. The large-scale behavior is determined by the closed periodic trajectories. The major oscillations observed in the spatial distribution of electron density and electrostatic potential are determined by the first classical periodic trajectories of an electron, namely the $(3,1)$ and $(4,1)$ orbits in a spherically symmetric potential well~\cite{Kuratov:2019,Kuratov:2021} as shown in Fig.~\ref{fig:hemisphere-model}a. Different possible effects related to macroscopic shell structure are considered in Refs.~\cite{Kuratov:2023,Kuratov:2024}. It has been shown that it leads to a modification of the spectrum of electronic oscillations in metal clusters, and also reveals a new mechanism of hydrodynamic cumulation driven by quantum shell effects.

In the present study, we discuss a novel evidence of the large-scale quantum shell effect in an uncharged isolated submicron hemispherical metal cluster, which consists in the existence of electrostatic field in the vicinity of its surface. This can be understood using the semiclassical Green's function method similarly to Refs.~\cite{roccia2010semiclassical,Kuratov:2019,Kuratov:2021} as follows. The closed periodic classical trajectories in a hemisphere are obtained similar to the ones in the sphere by reflection with respect to the flat part of the hemisphere as shown in Fig.~\ref{fig:hemisphere-model}a. This circumstance leads to the fact that the semiclassical theory applied for a sphere can be extended also to a hemisphere. In other words, near the flat surface there is an inhomogeneous distribution of charge, which leads to the appearance of an electrostatic field. Note that in the absence of this quantum effect, the charge density oscillations would be negligible, which corresponds to the classical case.

\begin{figure*}[t]
\centering
\includegraphics[width=0.99\textwidth]{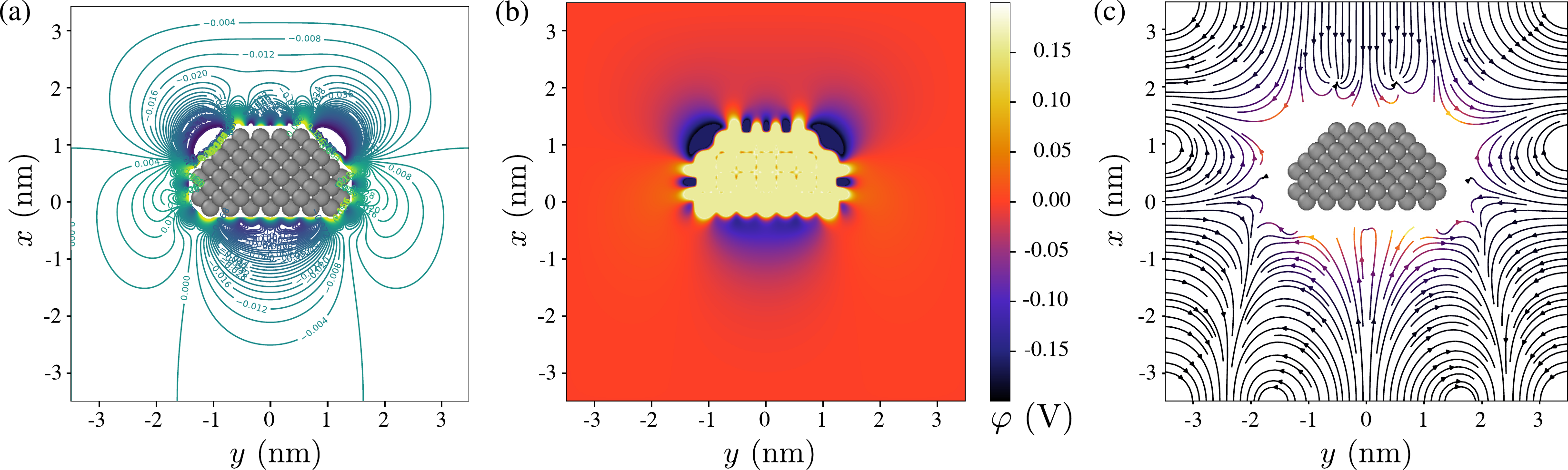}
\caption{Results for hemispherical Li cluster of $R = 1.4\,$nm composed of 288 atoms. (a) Contour plot for electrostatic potential values. (b) Electrostatic potential in $z = 0$ plane: colormap for the values higher than $0.2$~V caused by nuclear field is limited to $0.2$~V. (c) Electrostatic field stream lines in the exterior region of Li cluster. }
\label{fig:atoms_136}
\end{figure*}

To estimate the magnitude of the field theoretically, we use the results of the analysis performed for a sphere. The following dependencies has been obtained for the amplitude of the electron density disturbance inside the cluster in Refs.~\cite{Kuratov:2019,Kuratov:2021}:
\begin{equation}\label{eq:dne}
\Delta n_e \cong \frac{n_e}{\sqrt{N_e}}
\end{equation}
This allows us to estimate the magnitude of the charge oscillations inside the cluster with extrema in the cluster bulk
\begin{equation}\label{eq:dQ}
\Delta Q_e = \frac{1}{2} \Delta n_e e \frac{4\pi}{3}\left(\frac{R}{2} \right)^2 = \frac{1}{16} \frac{n_e}{\sqrt{N_e}} \frac{4\pi e}{3} R^3.
\end{equation}
This charge is redistributed inside the cluster bulk and on its surface. Based on the fact~\cite{Kuratov:2019,Kuratov:2021} that there are three extremum points in the distribution of electron density ($R_0/4, R_0/2, R_0/\sqrt{2}$), we approximate the charge distribution with five charged layers with the following radii and coordinates of their centers:
\begin{equation}
\label{eq:disk_field}
 \begin{aligned}
  & R_1 = R,             & & x_1 = 0,      \\
  & R_2 = \sqrt{15/16}R, & & x_2 = R/4,   \\
  & R_3 = \sqrt{3/4}R,   & & x_3 = R/2,   \\
  & R_4 = R/\sqrt{2},    & & x_4 = R/\sqrt{2},  \\
  & R_5 = R/4,           & & x_5 = \sqrt{15/16}R.
 \end{aligned}
\end{equation}
The layers have charges $-Q_1, Q_2, -Q_3, Q_4, -Q_5$ and surface areas $S_1 = \pi R^2, S_2 = 15/16 \pi R^2 , S_3 = 3/4 \pi R^2,  S_4 = 1/2 \pi R^2, S_5 = 1/16 \pi R^2$. To simplify the analysis, we assume that the charges are uniformly distributed over the disks with the same surface charge density $\sigma_1 = \sigma_3 = \sigma_5 = \sigma_{-},  \sigma_2 = \sigma_4 = \sigma_{+}$, as shown in Fig.~\ref{fig:hemisphere-model}b. The distribution of electric field provided by this ``layers model'' approach is shown in Fig.~\ref{fig:hemisphere-model}c.

Considering the electroneutrality of the cluster and Eq.~\eqref{eq:dne}, respectively, we have
\begin{equation}\label{eq:sigma}
\sigma_{+} = \frac{16 \Delta Q_e}{29 \pi R^2},\quad \sigma_{-} = \frac{16 \Delta Q_e}{23 \pi R^2}.
\end{equation}
Each i-th disk is a source of electrostatic field with the magnitude of
\begin{equation}
E_i = \frac{\sigma_i}{2 \epsilon_0} \left( 1 - \frac{r - x_i}{\sqrt{R_i^2 + (r - x_i)^2}} \right). \\  
\end{equation}

The obtained results provide a qualitative representation of the dependence of electrical quantities on the size of the cluster and the distance from it. Near the flat surface of the cluster ($r \ll R$), the electrostatic field decays linearly:
\begin{multline}\label{eq:linear_near_disk}
E = \sum_{i = 1}^5 E_i \approx \frac{\sigma_1}{2\epsilon_0} \left( 1 - \frac{r}{R} \right) + \frac{\sigma_2}{2\epsilon_0} \left( 1 - \frac{r + 0.25R}{R} \right) + ... \\ \approx \frac{\sigma_1 + 0.75 \sigma_2}{2\epsilon_0} - \frac{\sigma_1 + \sigma_2}{2\epsilon_0} \frac{r}{R} + ...
\end{multline}

At distances $r \gg R$, the field decays quadratically:
\begin{multline}\label{eq:quad_far_from_disk}
E = \sum_{i = 1}^5 E_i \approx \sum_{i = 1}^5 \left( 1 - \frac{r - x_i}{\sqrt{R_i^2 + (r - x_i)^2}} \right) \approx \\ \approx \sum_{i = 1}^5 \frac{\sigma_i}{2\epsilon_0} \frac{R_i^2}{2(r - x_i)^2}
\end{multline}

Further calculations are performed using the density functional theory (DFT) approach. In this case, the electrostatic potential of a system containing electrons and ions is given by: 
\begin{equation}\label{eq:potential}
V(\textbf{r}) = \sum_A \frac{Z_A}{|\textbf{R}_A - \textbf{r}|} - \int d\textbf{r}^\prime \frac{\rho(\textbf{r}^\prime)}{|\textbf{r}^\prime - \textbf{r}|},
\end{equation}
where the ionic contribution to electrostatic potential is represented by the first term which depends on the positions $\textbf{R}_A$ of the nuclei. The second term is determined by Hartree potential, which depends on electron density distribution $\rho(\textbf{r})$, 
calculated within \textit{ab initio} approach. There is a significant problem in the implementation of DFT modeling, which consists in the need to eliminate the influence of boundary conditions. A widely used numerical technique exploits plane-wave expansion of electron orbitals. This approach has become especially popular in pseudopotential codes (e.g., VASP~\cite{Kresse:1996}) and requires only one parameter, namely energy cutoff $E_\mathrm{cut}$, to be checked for convergence. However, this approach has a significant drawback: the unavoidable presence of periodic boundary conditions (PBC) imposed on computational domain edges~\cite{Stenlid:2019}. To decrease but not eliminate possible influence of PBC on calculated results, one has to artificially introduce enough empty volume around a nanocluster.

To obtain more reliable results for electrostatic field distribution around an isolated metallic hemisphere, we used GPAW \textit{ab initio} package~\cite{Enkovaara:2010}. It provides a finite-difference (FD) scheme for Kohn-Sham and Poisson equations, and the calculated results may be further mapped to a real-space grid. In this case, zero-boundary condition for electrostatic potential may be applied. Our choice of GPAW package is also caused by the provided support for PAW pseudopotential formalism for core states description~\cite{Blochl:1994}. We also used generalized gradient approximation (GGA) for exchange-correlation functional in the form of Perdew-Burke-Ernzerhof (PBE)~\cite{Perdew:1996}.

\begin{figure*}[t]
\centering
\includegraphics[width=0.99\textwidth]{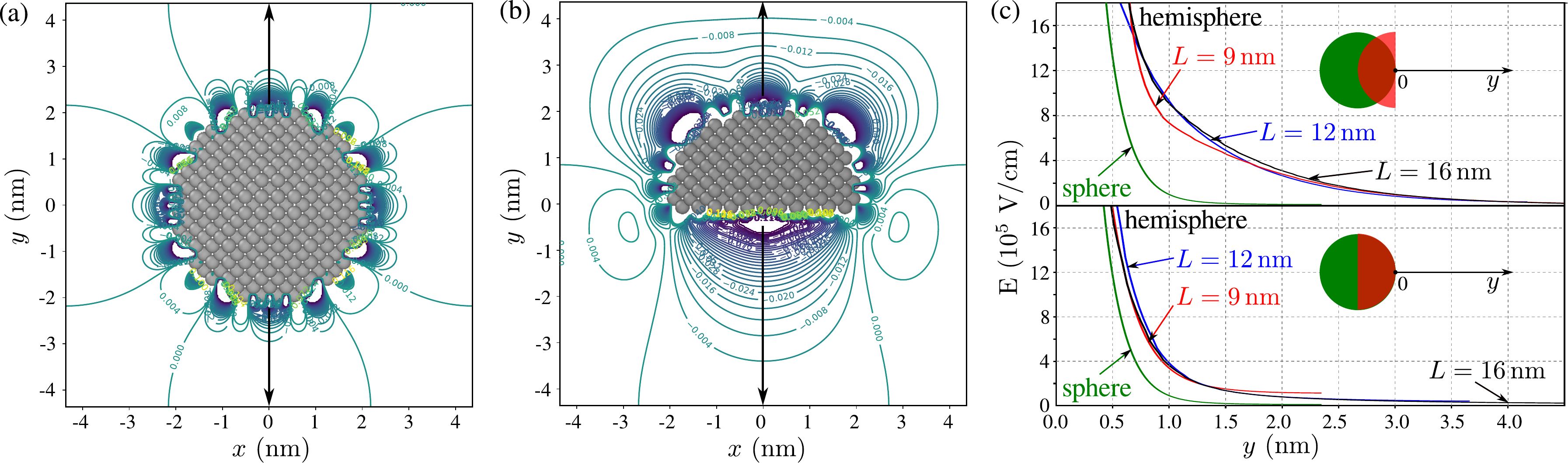}
\caption{Simulation results for spherical and hemispherical Li clusters of $R = 2\,$nm. (a) Contour plot for electrostatic potential values of spherical cluster. (b) Contour plot for electrostatic potential values of hemispherical cluster. (c) Electrostatic field magnitude along $y$-axis starting from the corresponding surface of a cluster: one may notice that boundary effects are negligible with domain size $L$ of 12 nm and larger.}
\label{fig:field_2nm}
\end{figure*}

Initial calculations are performed for a hemispherical cluster composed of $288$ Li atoms placed in the center of a $7 \times 7 \times 7$~nm$^3$ cubic box. The atoms were organized in a body-centered cubic (bcc) structure with a lattice parameter of $3.49$~\AA corresponding to normal material density and the radius of the modeled hemisphere is $1.4$~nm. The specificity of our calculations is that quite dense grids in real space are required. Even the most RAM-rich computers used in this work are unable to carry out computations on fine grids denser than $900 \times 900 \times 900$. The results for a $1.4$~nm hemisphere are obtained using the grid of $700 \times 700 \times 700$ cells (which means the cell size is less than $0.1$~\AA).

Besides the above described calculation, spherical and hemispherical lithium clusters with radii of $2$ and $2.5$~nm were considered, which contain significantly larger numbers of atoms: $960$ and $1600$, respectively. To control the influence of boundary conditions we additionally set the edge length $L$ of the cubic domain in simulations for hemispherical clusters to $9$, $12$, and $16$ nm. It is worth noting that the choice of lithium is driven by the possibility of using single-electron PAW pseudopotential in GPAW, so that the number of electrons in calculations is equal to number of ions, which allows to evaluate large clusters at less computational cost in a relatively large simulation domain.

The results for electrostatic potential distribution around a hemispherical Li cluster of $R=1.4$~nm are given in Fig.~\ref{fig:atoms_136}. The results shown in Fig.~\ref{fig:atoms_136}c provide the topology of the electrostatic field similar to that of an octupole. The obtained numerical results agree with the theoretical prediction in Fig.~\ref{fig:hemisphere-model}c. 

Applying the approximate model Eqs.~\eqref{eq:dQ}--\eqref{eq:disk_field} yields the following values for the surface density and electric field strength of a metallic cluster with radius $1$~nm and $n_e = 1.8 \cdot 10^{29}$~m$^{-3}$ near a flat surface
\begin{equation*}\label{eq:dQ_estim}
\Delta Q_e = 0.45 \times 10^{-18} Q,
\end{equation*}
\begin{equation*}\label{eq:sigma_estim}
\sigma_{+} = 0.104~Q/m^2, \sigma_{-} = 0.083~Q/m^2,
\end{equation*}
\begin{equation*}\label{eq:E_estim}
E \approx E_1 + E_2 + E_3 \approx \frac{\sigma_1 + 0.75 \sigma_2 + 0.5 \sigma_3}{2\epsilon_0} - \frac{\sigma_1 + \sigma_2 + \sigma_3}{2\epsilon_0} \frac{r}{R}.
\end{equation*}
Thus, the calculated field on a flat surface is $E \sim 10^7\,$V/cm, which is close in order of magnitude to the calculated value of $E\sim 2.0 \cdot 10^6\,$V/cm and decreases linearly for $r \le R$. We suppose that the difference between the calculated and theoretical values is explained by the fact that the theoretical model does not take into account the near-surface electron distribution spreading, which reduces the calculated electric field value. At distances greater than the cluster radius, the field decreases inversely proportional to the square of the distance, which also agrees with the results of theoretical analysis.

Fig.~\ref{fig:field_2nm} presents the results of numerical modeling for the electrostatic potential around spherical and hemispherical lithium clusters with radius $2$~nm. Fig.~\ref{fig:field_2nm}c shows an electrostatic field descent for these two cases. First one for the spherical cluster provides the field which acts predominantly at distances not exceeding $\sim 1\,$nm and has an order of magnitude smaller value than that for a hemisphere at the same distance. The presence of such a field around clusters of different shapes is a known physical phenomenon and is associated with the presence of an electronic ``aura'' around the cluster. This field is localized at distances not exceeding $1\,$nm for clusters of various sizes. It is easy to see that for the spherically symmetric case the calculated distribution of electron density at the surface of the sphere leads to smaller potential gradients than for a hemisphere of the same radius. 

A different character of the field behavior, caused by quantum shell effects, is observed in our study. The amplitude of the electrostatic field near flat surface the hemispherical cluster decreases with distance as follows
\begin{equation}\label{eq:inverse_root_law}
E \sim \frac{1}{\sqrt{R}}.
\end{equation}
One may notice, that the effect of boundary conditions is negligible for the cubic domain edges of $L = 12\,$nm and larger, so that such clusters may be considered as isolated ones. The result~\eqref{eq:inverse_root_law}, obtained from theoretical analysis, agrees with the numerical calculations presented in Fig.~\ref{fig:field_2nm}c: the field near the flat surface of the hemisphere acts on the scale of the radius and decreases in magnitude according to~Eq.~\eqref{eq:sigma}.

The performed theoretical and numerical analyses allow us to estimate the magnitude of the electrostatic field for large cluster sizes, for which simulations are not possible: 
\begin{itemize}
\item For a cluster size $R = 1$~nm at a distance of $0.5$~nm, $E = 10^6$~V/cm
\item For a cluster size $R = 10$~nm at a distance of $5$~nm, $E = 3.3 \cdot 10^5$~V/cm
\item For a cluster size $R = 100$~nm at a distance of $50$~nm, $E = 10^5$~V/cm
\end{itemize}

The analyzed effect can find numerous and effective applications in various areas of science and technology. 

\paragraph{Tweezers.}
The most obvious application of the effect under consideration is its use as tweezers for depositing molecules onto a flat surface of a cluster. Earlier different types of tweezers have been applied for individual molecules and nanoparticles manipulation, among which control mechanisms of magnetic~\cite{de2012recent} or optical~\cite{zhang2010trapping} nature are most widely used. In addition to the above mentioned techniques, some attention has been also attracted to electrostatic tweezers~\cite{fan2011electric,xu2016trapping}. In these works Coulomb force was manipulated in a special way by adjusting surface charge density with external voltage.

The results of our study allow for another approach for electrostatic tweezers construction, including Surface enhanced Raman spectroscopy (SERS) \cite{sharma2012sers}.

\paragraph{Nanoplasmonics.} 
Nanoparticles have been found to exhibit unique optical and electrical properties due to their small size and high surface-to-volume ratio~\cite{zamborini2012nanoparticles}. In particular, the local electric fields around NPs have been shown to play a crucial role in enhancing various physical, chemical and biological processes~\cite{estrada2012exploring}. These enhanced local electric fields can arise from the collective oscillation of free electrons at the nanoparticle surface, resulting in localized surface plasmon resonances (LSPRs)~\cite{chen2012review}. The specific characteristics of these LSPRs depend on factors such as the nanoparticle shape, size, and material composition.

The strong enhancement of local electric fields around nanoparticles has significant implications for various optical processes. SERS is a technique that exploits the enhanced local electric fields to amplify the Raman signal from molecules adsorbed on the nanoparticle surface. A system of hemispheres being a source of electrical polarization can be used to create SERS substrates that simultaneously exhibit pincers-like properties. Namely, local charges on the surface of the hemispheres create an electric field with a large spatial gradient, which should lead to the attraction of polarized molecules into the maximum of the field. As a result, an island film of such hemispheres can facilitate the deposition of molecules from solution.

\paragraph{Nanoelectronics.} 
One of the main modern trends in the development of semiconductor technology is miniaturization. At the present time, technologies have come closely to producing semiconductor elements of a nanometer scale~\cite{hossain2023advances}. The transition to structural elements of chips of this scale leads to the necessity to take into account effects that were previously insignificant on a larger scale. In particular, in the case of nanoscale semiconductor elements, it may be fundamentally necessary to take into account the quantum mechanical effects discussed in this paper. Moreover, it is possible to introduce new elements in the design of field-effect transistors. For example, a metallic element with a hemispherical shape on the surface of a semiconductor crystal can be used as a source of electric field which affects space charge and changes the local conductivity. Thus, it is possible to design a kind of passive gates in a semiconductor transistor to either lower or increase the threshold voltage.

Yet another application may include coating a system of hemispheres with a semiconductor that exhibits electro-optical properties, such as indium tin oxide (ITO)~\cite{ma2015indium}. Local areas around these hemispheres will demostrate different semiconductor properties as the effect of absorption in the optical range will occur. The hemispheres can be arranged in groups which sizes are of the order of the visible radiation wavelength. For ITO material to obtain metallic properties, a voltage of about $3$~V is required~\cite{ma2015indium}. Our DFT results show (Fig. \ref{fig:field_2nm}c) that the optical properties of ITO may be changed.

In general, the results of our first-principles DFT calculations demonstrate a novel effect which consists in the nonzero electric polarization of nanoparticles of hemispherical form, giving rise to an electric field within a large spatial domain. The emerged non-uniform distribution of volume and surface charge is attributed to the quantum shell effect previously described.

Significant electrostatic potential gradient around a submicron metallic hemisphere can serve for various applications. A system of such hemispheres can be used as electrostatic tweezers for mounting molecules onto a surface, allowing for precise manipulation of individual molecules. Additionally, the unique properties of hemispherical nanoparticles make them ideal candidates for SERS substrates that exhibit pincer-like properties, facilitating the deposition of molecules from solution. This promise great prospect for applications in biomedicine and chemical analysis.

The findings of this work also provide rather promising implications for nanoelectronics and optoelectronics. Metallic elements with hemispherical shapes can be used as sources of electric fields in field-effect transistors, enabling new design possibilities. Furthermore, coating a system of hemispheres with semiconducting materials that exhibit electro-optical properties can lead to local areas with different semiconductor properties. The arrangement of these hemispheres can be used to create optical effects in the range of light wavelengths, opening up new avenues for research in areas such as optoelectronics and photonics.

\begin{acknowledgments}
The authors gratefully thank Evgeny Andrianov, Alexander Baryshev, Alexander Dorofeenko, and Alexander Rakhmanov for valuable discussions and suggestions.
\end{acknowledgments}

\bibliography{library}

\end{document}